\documentclass[12pt]{article}

\usepackage{pdfpages}
\usepackage{eso-pic}
\usepackage{amssymb,amsmath,amsfonts}
\usepackage{url}
\usepackage{algorithm}
\usepackage[noend]{algpseudocode}

\def\rarrow#1{\hbox{$\xrightarrow{\hbox{$#1$}}$}}
\def\larrow#1{\hbox{$\xleftarrow{\hbox{$#1$}}$}}

\newcommand{\F}{\mathbb{F}}
\newcommand{\Sym}{\mathrm{Sym}}

\begin{document}

\title{On the Security of the Algebraic Eraser Tag Authentication Protocol}


\author{Simon R.~Blackburn\\
Information Security Group\\
Royal Holloway University of London\\
Egham, TW20 0EX, United Kingdom\\
and
\\
M.J.B.~Robshaw\\
Impinj\\
400 Fairview Ave. N., Suite 1200\\
Seattle, WA 98109, USA}

\maketitle

\begin{abstract}
The Algebraic Eraser has been gaining prominence as SecureRF, the company commercializing the algorithm, increases its marketing reach. The scheme is claimed to be well-suited to IoT applications but a lack of detail in available documentation has hampered peer-review. Recently more details of the system have emerged after a tag authentication protocol built using the Algebraic Eraser was proposed for standardization in ISO/IEC SC31 and SecureRF provided an open public description of the protocol. In this paper we describe a range of attacks on this protocol that include very efficient and practical tag impersonation as well as partial, and total, tag secret key recovery. Most of these results have been practically verified, they contrast with the 80-bit security that is claimed for the protocol, and they emphasize the importance of independent public review for any cryptographic proposal. 

\end{abstract}

\section{Introduction}

Extending security features to RAIN RFID tags\footnote{Following the creation of the RAIN Industry Alliance, 
UHF RFID tags are increasingly branded as RAIN RFID tags. These RFID tags operate at 860--960 MHz and are far more constrained
than the HF RFID tags that are familiar from public transport and NFC applications.} and other 
severely constrained devices in the {\it Internet of Things} is not easy. 
However the different pieces of the deployment puzzle are falling into place. Over-the-air (OTA) commands supporting security features have now been standardized~\cite{epcglobal-g2v2} and both tag and
reader manufacturers can build to these specifications knowing that interoperability will follow. The OTA commands themselves are crypto-agnostic so parallel work on a range of cryptographic
interfaces, so-called {\it cryptographic suites}, is ongoing within ISO/IEC SC31. These cryptographic suites provide the detailed specifications that allow algorithms such as the AES~\cite{nist-aes,29167-10}, 
PRESENT-80~\cite{present,29167-11}, and Grain-128a~\cite{grain128a,29167-13} to be used on even the most basic of RFID devices.

For symmetric cryptography a range of lighter alternatives to the Advanced Encryption Standard (AES)~\cite{nist-aes} have received a 
high level of cryptanalytic attention over several years. While the AES will always be an important implementation option, 
some of these alternative algorithms may be appropriate for certain use-cases. 
To those not in the field the cost and performance advantages provided by these new algorithms might appear slight. 
But the requirements of the RAIN RFID market are such 
that even a minor degradation in the read range or a small percentage increase in silicon 
price can eliminate the business case for adding security to many use-cases. 

Turning to asymmetric cryptography there are several work items in ISO/IEC 29167 that describe public-key solutions. Parts 29167-12~\cite{29167-12} and 29167-16~\cite{29167-16} describe tag authentication 
based on elliptic curve cryptography, 
though they carry significant implementation challenges for RAIN RFID. 29167-17~\cite{29167-17} 
provides another elliptic-curve tag authentication solution with the additional property that compact 
pre-computed coupons can be used to provide implementation advantages.
In 29167-20~\cite{29167-20}, however, we encounter an alternative to elliptic curves:
29167-20 proposes a method for asymmetric tag authentication that is based on \emph{braid groups}. This proposal 
is based on the \emph{Algebraic Eraser} (AE) key agreement protocol~\cite{ae-general,ae-protocol}.  SecureRF, the company commercializing (and owning the trademark to) the Algebraic Eraser, claims significant implementation advantages for the Algebraic Eraser over solutions that use elliptic curves. In particular the Algebraic Eraser is claimed to be well-suited to deployments as part of the Internet of Things.

\paragraph{Note}
The Algebraic Eraser has been proposed for use in many environments. However 
the commentary and descriptions in this paper will use the typical RFID setting of an Interrogator (or reader) interacting with a Tag. 
This provides the closest match with the terms used in the protocol~\cite{ae-protocol}.

\subsection*{Related work}

Until recently, crucial details about the Algebraic Eraser and any associated cryptographic protocol were not available. This made 
independent security analysis and performance evaluation difficult. (See~\cite{goldfeld,gunnells,kalka,myasnikov,securerf} for what 
little exists in the published literature.)
However, in October 2015 SecureRF provided a detailed public description of the Algebraic Eraser tag authentication 
protocol~\cite{stack-exchange,ae-protocol}. 
This means that the protocol can now be publicly reviewed and discussed. The published description includes a specific set of system 
parameters, a set of test vectors, and a description of the tag authentication protocol. 
However SecureRF do not disclose how the system parameters were generated, an aspect of the technology that is known to be
of crucial importance. Indeed, some of the attacks in this paper are able to 
exploit structure in the system parameters that have been proposed for standardization.

While general documentation~\cite{ae-general} describes the Algebraic Eraser in terms of braid groups, 
company presentations~\cite{ae-nist-paper,ae-nist-presentation} 
distance the technology from previous cryptographic proposals that use braid groups. Instead 
the security of the Algebraic Eraser is said to depend on a problem called the {\it simultaneous conjugacy separation search problem}~\cite{ae-nist-paper}
and sample parameter sizes have been published for different security levels. In~\cite{ae-protocol} the parameters 
are claimed to correspond to an 80-bit security level, though a precise security model is not provided. Most likely the intention is that the work effort to recover a private key from the corresponding public key should be roughly equivalent to $2^{80}$ operations. 

The tag authentication protocol in~\cite{ae-protocol} is based upon a Diffie--Hellman-like key agreement scheme. Very recently Ben-Zvi, 
Blackburn, and Tsaban~\cite{ben-zvi-etal} presented an innovative cryptanalysis of the underlying key agreement 
protocol.\footnote{The results in this paper are entirely independent of the work in Ben-Zvi \emph{et al}~\cite{ben-zvi-etal}.}  
Using only information that is exchanged over the air, and avoiding the hard problem upon which the security of the Algebraic Eraser 
key agreement protocol is claimed to be based, Ben-Zvi {\it et al.} provide a method for deriving the shared secret key. 
Using non-optimized implementations they successfully 
recovered---in under eight hours---the shared secrets generated using the Algebraic Eraser key agreement protocol with 
parameters provided by SecureRF that were intended to provide 128-bit security~\cite{ae-nist-presentation}. 

Since then Anshel, Atkins, Goldfeld and Gunnells (researchers associated with SecureRF) have posted a technical 
response~\cite{benzvi_rebuttal_arxiv} to the 
Ben-Zvi--Blackburn--Tsaban (BBT) attack. This is not the place to comment on that document, except
to highlight one feature that is relevant for our work here. 

In~\cite{benzvi_rebuttal_arxiv} Anshel {\it et al.}  
consider the implications of the BBT attack and 
state that the attack would not apply to one of two profiles proposed for standardization~\cite{29167-20}. 
Section~4.2 of Anshel {\it et al.}~\cite{benzvi_rebuttal_arxiv} reveals that the profile claimed to be secure 
is one where ``... an attacker never has access to one of the public keys ...''~\cite{benzvi_rebuttal_arxiv}. 
However the idea that it is reasonable for the security of a public key scheme to depend on the public key being hidden 
is very strange. 
While it is true Tag public keys could be delivered to interrogators out-of-band, the security 
of the scheme should not depend on the
Interrogator keeping those keys secret. Indeed, if we trust an Interrogator not to reveal the Tag public key then we can trust the Interrogator with a 
symmetric key and there would be no need to use the Algebraic Eraser at all! 
So while two of the five attacks described in this paper use the Tag public key for the required calculations, we see no limitation in assuming that the tag 
public key is, as the name implies, public.

Finally, we should point out a recent posting of Atkins and Goldfeld~\cite{AtkinsGoldfeld} that suggests modifications to the tag authentication protocol in the light of the results of this paper.
\subsection*{Our contribution}

In this work we derive a range of new and very efficient attacks on the tag authentication protocol~\cite{ae-protocol}. 
We side-step the bulk of the mathematical machinery behind the Algebraic Eraser, but observe some curious features of the Algebraic Eraser that cause significant failures in this protocol. In particular we provide the following attacks against the variant that is currently proposed for standardization:
\begin{enumerate}
\item Tag impersonation of a target tag with success probability $\approx2^{-7}$ after 273 queries against the target tag and storage $\approx 2^{16}$ bits.
\item Tag impersonation of a target tag with 100\% success rate after $\approx2^{15}$ queries against the target tag and using $\approx 2^{23}$ bits of storage.
\item Full recovery of a tag private key matrix (see Section~\ref{key_description}) with negligible work after running the tag authentication 
protocol 33 times against the target tag.
\item Tag impersonation of a target tag with 100\% success rate, using Attack~3 and a small pre-computed look-up table of around 128 64-bit words.
The on-line work in the attack is negligible while the off-line pre-computation for current parameter sizes is also negligible. This attack uses (a non-heuristic part of) an attack due to Kalka, Teicher and Tsaban~\cite{kalka} together with a novel application of a certain permutation group algorithm. 
\item Complete tag private key (or equivalent key) recovery---recovering both the tag private matrix and the secret tag conjugate 
set (see Section~\ref{key_description})---building 
on Attack~3 and requiring a work effort of $2^{49}$ operations and storage $\approx 2^{48}$ 64-bit words for one of the parameter 
choices proposed for standardization that is claimed to provide 80-bit security.  
\end{enumerate}
Our attacks avoid using any heuristic methods, and apply for all parameter sets of the size proposed in the standard (not just the specific given parameters). These failures in the tag authentication protocol severely undermine claims for an 80-bit security level. We conclude that the protocol is unsuitable for both deployment and standardization in its current form.

Our paper is structured as follows. In Section~\ref{sec:overview} we provide an overview of the Algebraic Eraser tag authentication protocol 
with the mathematical formalities following in Section~\ref{sec:technical}. The attacks are described in 
Sections~\ref{impersonation_attack},~\ref{leakage_attack},~\ref{RT_impersonation_attack}, and~\ref{full_leakage}  respectively and 
we close the paper with our conclusions.

\section{Algebraic Eraser and tag authentication}
\label{sec:overview}

\begin{table}[t]
$$\begin{array}{|ccc|}
\hline
{\textbf{Interrogator }} && {\textbf{Tag }} \\
\mbox{secret key } \mbox{\tt Int}_{\mbox{\small priv}}  &&  \mbox{secret key } \mbox{\tt Tag}_{\mbox{\small priv}} \\
\mbox{public key } \mbox{\tt Int}_{\mbox{\small pub}}   &&  \mbox{public key } \mbox{\tt Tag}_{\mbox{\small pub}}  \\
\hline
&&\\
\multicolumn{1}{|r}{\quad \mbox{Request Tag public key/}} & \rarrow{\quad  \mbox{\it start} \quad } & \\
\multicolumn{1}{|r}{\quad \mbox{certificate}} & & \\
&\larrow{\quad  \mbox{\tt Tag}_{\mbox{\small pub}} \quad } & \multicolumn{1}{l|}{\mbox{Send Tag public key} \>} \\
&& \\
\multicolumn{1}{|r}{\mbox{Send Interrogator public key, }} & \rarrow{\quad  \mbox{\tt Int}_{\mbox{\small pub}}, s, l \quad } & \\
\multicolumn{1}{|r}{\mbox{index, and token bit length}} & & \\
&& \\
 \multicolumn{1}{|r}{\quad\mbox{Compute secret using } \mbox{\tt Int}_{\mbox{\small priv}} \>}  && \multicolumn{1}{l|}{\mbox{Compute secret using } \mbox{\tt Tag}_{\mbox{\small priv}} \quad} \\
&& \\
\multicolumn{1}{|r}{\mbox{Check correctness of $t$}} & \larrow{\quad t \quad } & \multicolumn{1}{l|}{\mbox{Using index $s$ and length $l$}} \\
&& \multicolumn{1}{l|}{\mbox{extract and return token $t$ }} \\
&&\\
\hline
\end{array}$$ 
\caption{\label{tab:tag-auth}
An outline of the Algebraic Eraser tag authentication scheme~\cite{ae-protocol}. The underlying key agreement protocol is used to derive a shared secret. The Interrogator instructs the Tag, using byte index $s$ and bit-length $l$, to extract an 
authentication token $t$ of length $l$ from this shared secret.}
\end{table}

The Algebraic Eraser does not use familiar mathematics and a description can be, at first sight, somewhat complicated. 
However, for our attacks we will only need the basic tools that we provide in Section~\ref{sec:technical}.
For a more complete view the reader is referred to both the general description of the Algebraic Eraser~\cite{ae-general} and the specific protocol 
details in~\cite{ae-protocol}. 

As mentioned in the Introduction, at the core of the Algebraic Eraser is a key agreement protocol. Using the familiar protocol 
flow that dates back to Diffie--Hellman~\cite{diffie-hellman}, an Interrogator and Tag exchange public keys. Then,  by each applying their own secret component to the other public key, both Interrogator and Tag can arrive at a shared common secret key value. 
To turn this key agreement protocol into a tag authentication protocol,  the Interrogator specifies a portion of the shared secret that should be returned by the Tag. The correctness of this response can be verified by the Interrogator.
This is illustrated in Table~\ref{tab:tag-auth} and described more technically in Section~\ref{subsec:protocol}. 

We will refer to the portion of the secret key returned by the Tag as an authentication token $t$. In~\cite{ae-protocol} the Interrogator indicates to the Tag how to construct $t$ by sending a starting index $s$ and 
length $l$ during the message exchange between Interrogator and Tag.
The protocol description neither specifies nor gives guidelines on $s$ and $l$. Clearly a fake tag will always be able to fool an Interrogator with probability $2^{-l}$ but the 
field specifying the length $l$ in~\cite{ae-protocol} is eight bits long so we have $0\le l \le 255$. This certainly covers all the natural choices. 
Note that generating an authentication token $t$ by revealing parts of a shared secret means that the Interrogator will need to generate and use different public keys at each tag authentication. While 
this is alluded to in Section B.1.2 of~\cite{ae-protocol} it is unclear whether the ensuing performance penalty in storage and transaction time is always reflected in published performance figures.

\section{Some technical details}
\label{sec:technical}

This section reviews some of the technical details of the protocol. We describe only as much of the detail as we need to describe our attacks.

\subsection{System parameters}

The protocol specifies some system parameters, the \emph{key space}, as follows.

Let $N$ be a small positive integer;~\cite{ae-protocol} mandates that $N=10$. 
Let $B=\{b_1,b_2,\ldots ,b_{N-1}\}$ be an alphabet of size $N-1$ (the $b_i$ are known as \emph{Artin generators}). 
Let $F$ be the set of all formal strings in the disjoint union $B\cup B^{-1}$. So, for example, $b_2b_1^{-1}b_1b_4b_2^{-1}$ is a length 5 element of $F$. 

The \emph{set of Tag conjugates} is a set $C=\{c_1,c_2,\ldots ,c_{32}\}$ of size 32, where each $c_i\in F$. 
The \emph{set of Interrogator conjugates} is a set 
$D=\{d_1,d_2,\ldots ,d_{32}\}$ of size 32, where each $d_i\in F$. While $C$ and $D$ are specified in~\cite{ae-protocol}
SecureRF does not describe how they
have been generated. In fact, in Section~\ref{impersonation_attack} we will exploit an important structural property of 
the sets $C$ and $D$ that have been proposed for standardization. Here, however, we restrict ourselves to noting that each of $C$ and $D$ require around 90 Kbits to store and that
while a tag might not need to store $C$ the Interrogator needs $D$ to generate ephemeral keys.

We write $\Sym(N)$ for the set of all permutations of $N$ objects $\{1,2,\ldots ,N\}$. 
Let $s_i=(i, i+1)\in\Sym(N)$ be the permutation that swaps $i$ and $i+1$ and leaves the remaining elements fixed. Let
\[
w=b_{i_1}^{\epsilon_1}b_{i_2}^{\epsilon_2}\cdots b_{i_r}^{\epsilon_r}
\]
be an element in $F$ of length $r$, where $i_j\in\{1,2,\ldots ,N-1\}$ and $\epsilon_j\in\{-1,1\}$. 
The \emph{permutation $\pi(w)\in\Sym(N)$ corresponding to $w\in F$} is the permutation
\[
\pi(w)=s_{i_1}^{\epsilon_1}s_{i_2}^{\epsilon_2}\cdots s_{i_r}^{\epsilon_r}=s_{i_1}s_{i_2}\cdots s_{i_r}
\]
where product means composition of permutations.

Finally, the protocol~\cite{ae-protocol} specifies using arithmetic in the finite field $\F_{256}$ and defines a specific sequence of $N=10$ non-zero elements in $\F_{256}$, 
called \emph{T-values}, and a specific $N\times N$ matrix $M_*$ with entries in $\F_{256}$ called a \emph{seed matrix}. 
This choice of parameter sizes is denoted B10F256 and, according to Section B.3, is intended to provide 80-bit security. 

Another set of parameters, denoted B16F256, has been independently provided by SecureRF 
to the first author. The same underlying field is used for both parameter sets but the matrices, 
the set of T-values, and the permutations are defined for $N=16$ rather than $N=10$. The parameters B16F256 are 
intended to provide 128-bit security.

\subsection{E-multiplication}
\label{mult-desc}

E-multiplication is the public key operation, analogous to finite field exponentiation in Diffie--Hellmann, that lies 
at the heart of the Algebraic Eraser. It takes two parameters as input. 
The first parameter is a pair $(M,\sigma)$ where $M$ is an $N\times N$ matrix over $\F_{256}$ and $\sigma\in\Sym(N)$ is a permutation. The second parameter is a string $w\in F$. The 
output is a pair $(M',\sigma')$ where $M'$ is an $N\times N$ matrix over $\F_{256}$ and $\sigma'\in\Sym(N)$. 
We write
\[
(M,\sigma)*w=(M',\sigma').
\]
The permutation $\sigma'$ is easy to define: $\sigma'=\sigma\, \pi(w)$. The matrix $M'$ is computed by first finding a certain $N\times N$ matrix $\phi(\sigma,w)$ with entries in $\F_{256}$, and then 
setting $M'=M \phi(\sigma,w)$. We do not specify the details of how $\phi(\sigma,w)$ is defined, but  
just give the following details. To compute $\phi(\sigma,w)$, we replace the symbols $b_i$ and $b_i^{-1}$ in $w$ by certain fixed matrices and their inverses. These matrices have entries in a polynomial ring in $N$ variables, and the last row of all these matrices is all zero apart from the final entry which is~$1$. We multiply our matrices together (obtaining a matrix whose last row is all zero apart from the final entry which is~$1$). We evaluate each entry of this product (which is a ratio of two polynomials in $N$ variables) by replacing each variable by one of the $T$-values to form the matrix $\phi(\sigma,w)$ with entries in $\F_{256}$. We use $\sigma$ to decide which $T$-value replaces each variable in this process.

We note four properties that follow from the way E-multiplication is defined:

\begin{enumerate}
\item If $w$ is the concatenation of strings $w'$ and $w''$ then
\begin{equation}
\label{eqn:concatenation}
(M,\sigma)*w=((M,\sigma)*w')*w''.
\end{equation}
In fact E-multiplication has other nice properties related to the fact that E-multiplication is derived from the action of a braid group. 
However we do not need these properties here.
\item The matrix $\phi(\sigma,w)$ only depends on $\sigma$ and $w$ (and on the T-values, which are fixed).
\item The entries of the last row of $\phi(\sigma,w)$ are all zero, except the final entry which is $1$. 
\item The following linearity property follows from our partial description of E-multiplication:
\begin{equation}
\label{eqn:linearity}
\begin{split}
\text{If }(M_1,\sigma)*w=(M'_1,\sigma')\text{ and }(M_2,\sigma)*w=(M'_2,\sigma')\\
\text{ then }
(a_1M_1\oplus a_2M_2,\sigma)*w=(a_1M'_1\oplus a_2M'_2,\sigma')
\end{split}
\end{equation}
for any $a_1,a_2\in\F_{256}$.
\end{enumerate}

\subsection{Private and public keys}
\label{key_description}

The Tag private key has two components. 
\begin{enumerate}
\item The first component is an $N\times N$ matrix $K_T$ over $\F_{256}$ that is generated from the seed matrix $M_*$. 
During the key generation process a random degree 9 polynomial $p(x)$ over $\F_{256}$ is selected and we set $K_T= p(M_*)$.
See Section B.1.2 of ~\cite{ae-protocol}. The parameters are chosen so that the probability of recovering $K_T$ by guessing the polynomial $p(x)$ is 
$(2^{-8})^{10} = 2^{-80}$. 
\item The second component of the private key is a string $c\in F$ that is obtained by concatenating at least 16 of the 
Tag conjugates and their inverses. (The inverse of a word $b_{i_1}^{\epsilon_{1}}b_{i_2}^{\epsilon_{2}}\cdots b_{i_r}^{\epsilon_{r}}$ 
is the word $b_{i_r}^{-\epsilon_{r}}b_{i_{r-1}}^{-\epsilon_{r-1}}\cdots b_{i_1}^{-\epsilon_{1}}$.)
\end{enumerate}
The 
matrix $K_T$ and the string $c$ form the \emph{private key} of the Tag. The Tag's \emph{public key} is defined to be
\[
(M_T,\sigma_T)=(K_T,1)*c
\]
where $1$ is the identity permutation.

When interacting with the Tag, the Interrogator generates an ephemeral private and public key, using the set of Interrogator conjugates 
rather than Tag conjugates. 
This means that the Interrogator's private key is an $N\times N$ matrix $K_I$ over $\F_{256}$ and a concatenation $d$ of at least 16 of the Interrogator conjugates and their inverses. 
The Interrogator's public key is
\[
(M_I,\sigma_I)=(K_I,1)*d.
\]

\subsection{Authenticating a tag}
\label{subsec:protocol}

The Tag authentication protocol runs as follows. The Interrogator requests the Tag's public key $(M_T,\sigma_T)$. The Interrogator 
also generates an 
ephemeral private key and sends the corresponding public key $(M_I,\sigma_I)$ to the Tag. The Tag computes the shared key
\[
(K_TM_I,\sigma_I)*c
\]
and the Interrogator computes the shared key
\[
(K_IM_T,\sigma_T)*d.
\]
The function $\phi$ and the parameters of the scheme are designed so that these values are equal.
The Interrogator requests that part of the shared key be returned to the Interrogator and authenticates the Tag if the Tag replies correctly. Though the shared key is a matrix-permutation pair, the permutation is easy to compute from public material (it is just a product of two public permutations: $\sigma_I\sigma_T=\sigma_T\sigma_I$). So the matrix is the only non-public part of the shared key.

We note that all the attacks in this paper use knowledge of the shared secret key generated during the tag authentication protocol. 
It is a minor detail, but since~\cite{ae-protocol} restricts the length of the authentication token ($l\le255$) an attacker might need to repeat tag authentication 
using three different choices for $s$ and $l=255$ before recovering the entire shared secret (as the shared matrix is represented by a sequence of $8\times N(N-1)=720$ bits). This three-fold 
increase in the work effort is included in our estimates. 

\section{Basic tag impersonation}
\label{impersonation_attack}

In a tag authentication protocol, an attacker can always run the tag authentication protocol 
against a target tag at will. The goal would be to derive enough information so that the attacker can impersonate the target tag to a genuine Interrogator in a future run of the tag authentication protocol. We now describe a simple impersonation attack of this type.

Suppose an attacker chooses a permutation $\sigma$ and a set of matrices $M_i$, for $0\le i\le N(N-1)=90$. The matrices 
are chosen so that they form a basis
for the space of all $N \times N$ $\F_{256}$ matrices for which the last row begins with $N-1$ zero values. Taken together, the matrices and the single permutation
$\sigma$ provide $N(N-1)+1=91$ spoof Interrogator public keys that are used in 91 runs of the tag authentication protocol against the target Tag. This yields 91 
shared secrets $S_i$, for $0\le i\le N(N-1)$, remembering from Section~\ref{subsec:protocol} that we will need to include a further factor of three in any work effort computation.

Now suppose the attacker attempts to impersonate the target Tag to a genuine Interrogator and receives a random public key $(M_I, \sigma_I)$, where $\sigma_I=\sigma$. Emulating the target Tag, the attacker computes $a_i$ for $0\le i\le N(N-1)$, so that
$$ M_I = \bigoplus_{i=0}^{N(N-1)} a_i M_i.$$
The linearity observed in Equation~\ref{eqn:linearity} of Section~\ref{mult-desc} 
guarantees that the secret $S$ that would be computed by a genuine tag can also be computed as
$$ S = \bigoplus_{i=0}^{N(N-1)} a_i S_i.$$
The attacker will be able to extract the correct authentication token from $S$ and fool the Interrogator with 100\% certainty.
 
As described, the attack requires that the Interrogator choose a public key with $\sigma_I=\sigma$. At first sight, for 
the parameters in~\cite{ae-protocol}, it appears that since $N! \approx 2^{21.8}$ the 
probability a genuine Interrogator chooses the hoped-for $\sigma_I$ is around $2^{-21.8}$.  However closer analysis reveals additional structure
in the conjugate sets $C$ and $D$. In particular, all the permutations generated by $C$ have
five fixed points, as do all the permutations generated by $D$. This means that the space of possible permutations that
might be encountered from a genuine Interrogator is reduced from $N!$ to $(N/2)! \approx 2^{7}$. The 
probability a genuine Interrogator chooses the hoped-for $\sigma_I$ is therefore greater than $2^{-7}$.

For those that prefer certainty, it is obvious an attacker can increase his success probability by performing more off-line interrogation of the 
target Tag using different $\sigma$. This gives a variety of trade-offs, with the extreme being an 
attacker who will be able to emulate the target tag with 100\% certainty after interrogating that tag 
around $91 \times 3 \times 5! < 2^{15}$ times. 

\section{Tag private matrix recovery}
\label{leakage_attack}

The security of the Algebraic Eraser tag authentication protocol depends on the secrecy of two components: the $N \times N$ 
private $\F_{256}$-matrix $K_T$ and the tag string $c \in F$. In fact, both of these need to be kept secret: in the section below we provide details of a very efficient tag impersonation attack if $K_T$ is known; moreover, $K_T$ can be recovered from the public key if $c$ is known.
In Section B.3 of~\cite{ae-protocol} parameters are chosen so that the work effort to recover $K_T$ by guessing the polynomial $p(x)$ used to construct it is equal to the claimed security level of $2^{80}$ operations. 

Exploiting the linearity observed in Equation~\ref{eqn:linearity} of Section~\ref{mult-desc} we show how a differential cryptanalytic attack can recover 
the entirety of the secret matrix $K_T$ after 11 tag authentications. Taking into account protocol constraints and parameters 
specified in~\cite{ae-protocol} we will need 33 tag authentications in practice, but in the following description 
we will set aside the factor of three for clarity.

To begin, the attacker authenticates a target Tag using any Interrogator public key $(A, \sigma)$ and stores the shared secret $S$ that results.  
The attacker then authenticates the same tag with N related public keys that use the same permutation $\sigma$ and 
matrices $P_1, \ldots, P_{N}$ constructed as follows. 

Let $E_{i,j}$ be the $N\times N$ matrix that is all zero, except its $(i,j)$ entry which is $1$.
Set $P_t = A\oplus E_{t,N}$ for $1\le t \le N$.
The attacker challenges the target tag with the ten public keys  $(P_t, \sigma)$, for $1 \le t \le N$, and stores the secret matrices $S_t$ that result.

One can observe that $S = K_T A V$ and $S_t = K_T  P_t  V$,  for $1 \le t \le N$, where the matrix
$V=\phi(\sigma,c)$ will depend on $\sigma$ and the Tag's secret product $c$ in a complicated way; the last row of $V$ is all zero, except its last entry which is 1, by a property of E-multiplication stated above. However neither $P_t$ nor $V$ depend on $K_T$ and we observe that 
$$ S \oplus S_t = (K_T  A  V) \oplus (K_T  P_t V) =  K_T (A \oplus P_t) V=K_TE_{t,N}V.$$
Since the last row of $V$ has a special form, 
$S\oplus S_t$ will be zero everywhere except in the last column, for $1 \le t \le N$. 
Further, the values in this last column will correspond to the $t^{\rm th}$ column of the tag secret matrix $K_T$.  Taken together, 
the entirety of the tag secret matrix $K_T$ can be recovered column-by-column and 
something that is intended to require $2^{80}$ operations can be accomplished with negligible work after $N+1=11$ interactions with the target Tag, or 
33 tag authentications if we take into account the protocol constraints in~\cite{ae-protocol}.  

This attack has been confirmed using the parameters and examples given in~\cite{ae-protocol}. It has also been confirmed on 
parameter sets of the form B16F256 that have been supplied by SecureRF. In this latter case, with N=16, we are required to perform 17 
interactions with the target Tag, or 136 tag authentications if we respect protocol considerations and only recover at most 255 bits in each interaction. Recall that parameter sets of the form
B16F256 are intended to provide 128-bit security. 

The linearity property that facilitates this attack appears intrinsic to the definition of the Algebraic Eraser and thus hard to avoid; 
increasing the size of parameters will not provide any significant additional security.

\section{Efficient tag impersonation}
\label{RT_impersonation_attack}

Even though the tag impersonation attack of Section~\ref{impersonation_attack} is already very effective, a more efficient 
attack can be designed using the result of Section~\ref{leakage_attack}. This new attack is more efficient in terms of all three measures of tag 
queries, computation, and storage.

Recall that $d_1,d_2,\ldots,d_{32}\in F$ are the interrogator conjugates. Define their corresponding permutations $g_i\in \Sym(N)$ by $g_i=\pi(d_i)$.
We already observed in Section~\ref{impersonation_attack} that these permutations are highly structured and have five fixed points. 

Stage~0: A pre-computation stage. Build an oracle which, when given a permutation $\sigma\in\Sym(N)$ that lies in the 
subgroup of $\Sym(N)$ generated by the $g_i$, returns $r$ (a small integer), $i_1,i_2,\ldots,i_r\in\{1,2,\ldots ,32\}$ and 
$\epsilon_1,\ldots ,\epsilon_r\in\{-1,1\}$ such that
\[
\sigma=g_{i_1}^{\epsilon_1}g_{i_2}^{\epsilon_2}\cdots g_{i_r}^{\epsilon_r}.
\]

Since $N!=10!\leq 2^{22}$, we can build a very efficient oracle by constructing a lookup table of size $N!$ which contains the pair $i_r$ and $\epsilon_r$ for each permutation~$\sigma$ that can be written as a product of the $g_i$ (and a termination string for the identity permutation). The table may be constructed by using Algorithm~\ref{lookup_algorithm}.

\begin{algorithm}[t]
\caption{Constructing a lookup table}
\label{lookup_algorithm}
\begin{algorithmic}[1]
\State Construct a table indexed by the $N!$ permutations in $\Sym(N)$, with all entries empty.
\State Add `terminate' to the entry corresponding to the identity permutation.
\State Let $L$ be a list that contains just the identity permutation.
\While{$L$ non-empty}
\State Let $g$ be the first element in $L$.
\For{$i\in\{1,2,\ldots,32\}$ and $e\in\{-1,1\}$}
\State Compute $gg_i^e$.
\If{the table entry indexed by $gg_i^e$ is still empty}
\State Change this entry to $(i,e)$.
\State Add $gg_i^e$ to $L$.
\EndIf
\EndFor
\State Remove $g$ from $L$.
\EndWhile
\end{algorithmic}
\end{algorithm}

Since each permutation $g$ is added to the list $L$ at most once, constructing the table takes at most about $N!\times 32\times 2\approx 2^{28}$ operations. Once the table is constructed, the oracle works on input $\sigma$ by using the table to find the last element in a product of the permutations $g_i$ and their inverses that is equal to~$\sigma$. It then multiplies $\sigma$ by the inverse of this last element, and iterates until it reaches the identity permutation. The oracle returns the shortest expression of the form we want (though we do not need this). The oracle is very efficient: just a few table lookups and permutation compositions are needed.

The subgroup generated by the permutations $g_i$ in~\cite{ae-protocol} is extremely small (as these permutations all fix the same five points). So building the table for the oracle above is extremely fast. We have implemented Algorithm~\ref{lookup_algorithm} in C. It takes 
just 0.014 seconds to generate the table, and resulting oracle takes an average of under 0.00005 seconds to answer typical query, running on a 2.7GHz i7 MacBook Pro. So the pre-computation stage takes a negligible time to complete, and the resulting oracle is extremely fast in practice.

Note that Algorithm~1 and the resulting oracle are very efficient even if the permutations $d_i$ generate the whole of the symmetric group (the worst case for the pre-computation). Experiments with our implementation show that the table is constructed in 66 seconds, and the resulting oracle answers a typical query in an average of under 0.0015 seconds. So the pre-computation is always efficient.

For situations where it becomes impossible to store (in RAM) a table of length equal to the order of the subgroup generated by the permutations $g_i$, for example if $N$ is much larger, we would suggest first using standard Schreier--Sims techniques (see Seress~\cite[Chapter~4]{Seress}, for example) and then the powerful heuristic approach of Kalka, Teicher and Tsaban~\cite{kalka}, to construct the oracle.  Note that the pre-computed oracle can be used whenever the same set of reader conjugates are used. Since the reader conjugate set $D$ is a public system parameter~\cite{ae-protocol} an oracle can be collaboratively computed and shared over the Internet. 

Stage 1: Interact with the Tag as in Section~\ref{leakage_attack} to obtain the Tag's public key $(M_T,\sigma_T)$ and then its secret key $K_T$.

Stage 2: Impersonate the Tag using the techniques of Phase 2 of the attack of Kalka, Teicher and Tsaban~\cite[Section~3.2.2]{kalka}. The details are as follows. 

When a legitimate interrogator queries $(M_I,\sigma_I)$, query the oracle to obtain
$i_1,i_2,\ldots,i_r\in\{1,2,\ldots ,32\}$ and $\epsilon_1,\ldots ,\epsilon_r\in\{-1,1\}$ such that
\[
\sigma_I=g_{i_1}^{\epsilon_1}g_{i_2}^{\epsilon_2}\cdots g_{i_r}^{\epsilon_r}.
\]
Define
\[
w=d_{i_1}^{\epsilon_1}d_{i_2}^{\epsilon_2}\cdots d_{i_r}^{\epsilon_r}.
\]
Compute the matrix $L_1$ that is the result of the following E-multiplication:
\[
(K_TM_I,\sigma_I)*w^{-1}.
\]
Compute the matrix $L_2$ that is the result of the following E-multiplication:
\[
(K_T^{-1}M_T,\sigma_T)*w.
\]
The shared key is $(L_1L_2,\sigma_T\sigma_I)$. This derivation has been implemented and confirmed.

\section{Full private key recovery}
\label{full_leakage}

Given the extreme effectiveness of the tag impersonation attack of Section~\ref{RT_impersonation_attack} the need for a full key recovery
attack on the Algebraic Eraser tag authentication protocol is questionable. Under normal circumstances one might prefer a key-recovery attack 
so that recovered keys could be inserted into a cloned device, thereby exploiting the storage and performance advantages of the original
algorithm. However, in our attacks, the pre-computed look-up table is small and impersonation is exceptionally fast; in fact 
it would be interesting to compare the performance of the impersonation attack to the computation required by the legitimate tag.

Nevertheless, to illustrate that a complete key recovery attack does exist we outline a basic attack using a {\it meet-in-the-middle} technique. While
the attack in this section is already very effective ($2^{48}$ storage and $2^{49}$ time for one of the parameter choices proposed for standardization)  
we believe that more analysis could reveal more practical variants. 

To start, we will say that Tag conjugate products $c,c'\in F$ are \emph{equivalent}, which we write as $c\equiv c'$, if
\[
(I,1)*c=(I,1)*c'
\]
where $I$ is the $N\times N$ identity matrix and where $1$ is the identity permutation. The definition of E-multiplication 
shows that when $V$ is any fixed invertible matrix  $c\equiv c'$ if and only if
\[
(V,1)*c=(V,1)*c'.
\]
In particular, when $V = K_T$ and $c$ are the two components of the Tag private key, a private key consisting of $K_T$ and $c'$ will produce the same Tag public key if, and only if, $c\equiv c'$ (because $K_T$ is invertible). Since all 
shared keys can be derived from the public key and the interrogator's secret information, replacing $c$ by $c'$ in the Tag makes no difference to any of the shared keys computed by the Tag in the protocol. So to recover the full secret key of the Tag we need only find $c'\in F$ that is equivalent to $c$. 

Assume that the Tag's secret product $c$ of conjugates has length $16$, as allowed by~\cite{ae-protocol}. There are $2\times 32=2^6$ possibilities for each term in the product, and so there are $2^{6\times 16}=2^{96}$ possibilities for $c$. We now 
describe a simple meet-in-the-middle technique that will recover an equivalent product $c'$ using a look-up table with $\sqrt{2^{96}}=2^{48}$ entries. The attack extends in a natural way to longer products of conjugates.

Suppose now that an attacker has recovered the Tag private matrix $K_T$ by the attack of Section~\ref{leakage_attack}. Clearly the attacker has the Tag's public key $(M_T,\sigma_T)$. The attacker then searches for 
products $c'\in F$ of Tag conjugates that are equivalent to $c$ by finding $c'$ such that $(K_T,1)*c'=(M_T,\sigma_T)$. We write $c'=w_1'(w_2')^{-1}$ where the $w_i$ are length eight products of Tag conjugates and their inverses. Note that
\[
(K_T,1)*w_1'=(M_T,\sigma_T)*w_2'.
\]
For each of the $2^{48}$ possibilities for $w'_1$, we compute $(K_T,1)*w_1'$. We store the results in such a way that it is easy to find $w_1'$ if we are given $(K_T,1)*w_1'$. For example, we could use an array of pairs $((K_T,1)*w_1',w_1')$, sorted by its first component. 

For each of the $2^{48}$ possibilities for $w'_2$, we compute  $(M_T,\sigma_T)*w_2$ and check whether this value occurs as the first of a pair in our array. Once we find such a value $w_2$, we set $c'=w_1'(w'_2)^{-1}$ where $w_1'$ is the second element of the pair we have found in the array. Note that
\[
(K_T,1)*c'=((K_T,1)*w'_1)*(w_2')^{-1}=((M_T,\sigma_T)*w_2')*(w_2')^{-1}=(M_T,\sigma_T),
\]
and so $c'$ and $K_T$ form a private key that produces the Tag's public key. Hence $c\equiv c'$, and we have found an equivalent private key for the Tag.  

Small-scale variants of this attack---using a reduced Tag conjugate set $C$ and shorter products---have been successfully implemented for the parameter sets B10F256 given in~\cite{ae-protocol}.

\section{Conclusion}
\label{further_comments}

The Algebraic Eraser has been on the periphery of the cryptographic literature for nearly ten years. 
However the designers have not made it easy for independent researchers to analyze the scheme. 
The reason for this approach is unclear, but the consequence has been a lack of independent peer-review. 

It is too soon to determine whether or not secure schemes can be built around the mechanisms seen in the Algebraic Eraser. 
Certainly it is always interesting to see new techniques based on different hard problems. 
But any performance claims for the Algebraic Eraser are premature without a more 
complete understanding of the security that is delivered. The work of Ben-Zvi {\it et al}~\cite{ben-zvi-etal} and that presented in this paper 
suggest that a lack of independent analysis has hindered the algorithm proponents from seeking out alternative viewpoints and, critically, 
from recognizing some very effective attacks. These have only become apparent 
as the profile of the algorithm has been raised and details about the algorithm have been made public.

It is hard to avoid the conclusion that the Algebraic Eraser should not be used or standardized in 
its current form. If future versions are proposed, and~\cite{AtkinsGoldfeld} 
provides hints that this may be the case, then it is important that a full and detailed specification be made publicly available. Just as for the parent algorithm, we believe any variants should not be used until there has been sufficient independent public cryptanalysis.

\end{document}